\documentclass[twocolumn,aps,prl,superscriptaddress]{revtex4-1} 
\usepackage{graphicx}
\usepackage{epstopdf}
\usepackage{extarrows}
\usepackage{textcomp}
\usepackage{fullpage}
\usepackage{pslatex}
\usepackage{amsmath}
\usepackage[left=20mm, right=20mm, top=15mm, bottom=20mm]{geometry}

\begin{document}

\title{Career on the Move: Geography, Stratification, and Scientific Impact}

\author{Pierre Deville}
\affiliation{\footnotesize{Department of Applied Mathematics, Universit\'e catholique de Louvain, Belgium}}
\affiliation{\footnotesize{CCNR and Physics Department, Northeastern University, Boston, MA 02115, USA}}
\author{Dashun Wang}
\affiliation{\footnotesize IBM Thomas J. Watson Research Center, Yorktown Heights, NY, 10598, USA}
\affiliation{\footnotesize{CCNR and Physics Department, Northeastern University, Boston, MA 02115, USA}}
\author{Roberta Sinatra}
\affiliation{\footnotesize{CCNR and Physics Department, Northeastern University, Boston, MA 02115, USA}}
\affiliation{\footnotesize{Center for Cancer Systems Biology, Dana-Farber Cancer Institute, Boston, MA 02115, USA}}
\author{Chaoming Song}
\affiliation{\footnotesize Department of Physics, University of Miami, Coral Gables, FL 33124,USA}
\affiliation{\footnotesize{CCNR and Physics Department, Northeastern University, Boston, MA 02115, USA}}
\author{Vincent D. Blondel}
\affiliation{\footnotesize{Department of Applied Mathematics, Universit\'e catholique de Louvain, Belgium}}
\author{Albert-L\'aszl\'o Barab\'asi $^{\ast,}$}
\affiliation{\footnotesize{CCNR and Physics Department, Northeastern University, Boston, MA 02115, USA}}
\affiliation{\footnotesize{Department of Medicine, Brigham and Women's Hospital, Harvard Medical School, Boston, MA 02115}}
\affiliation{\footnotesize Center for Network Science, Central European University, Budapest, Hungary,\\ $^\ast$ alb@neu.edu}

\begin{abstract}

Changing institutions is an integral part of an academic life. Yet little is known about the mobility patterns of scientists at an institutional level and how these career choices affect scientific outcomes. Here, we examine over 420,000 papers, to track the affiliation information of individual scientists, allowing us to reconstruct their career trajectories over decades. We find that career movements are not only temporally and spatially localized, but also characterized by a high degree of stratification in institutional ranking. When cross-group movement occurs, we find that while going from elite to lower-rank institutions on average associates with modest decrease in scientific performance, transitioning into elite institutions does not result in subsequent  performance gain. These results offer empirical evidence on institutional level career choices and movements and have potential implications for science policy. 

\end{abstract}

\maketitle
\newpage

Despite their importance for education, scientific productivity, reward and hiring procedures, our quantitative understandings of how individuals make career moves and relocate to new institutions, and how such moves shape and affect performance, remains limited. Indeed, previous research on migration patterns of scientists \cite{auriol2007labour,auriol2010careers}  tended to focus on large-scale surveys on country-level movements, revealing long-term cultural and economical priorities \cite{vannoorden2004,schiermeier2011career,jans2010study, solimano2008international}. At a much finer scale, research on human dynamics and mobility has emerged as an active line of enquiry  \cite{brockmann/nature/2006,simini2012universal,PhysRevLett.86.3200,gonzalez/nature/2008,bagrow2011collective, lu2012predictability,szell2012understanding}, owing to new and increasingly available massive datasets providing time resolved individual trajectories \cite{blondel2012data}. While these studies cover a much shorter time scale than a typical career, they uncover a set of regularities and reproducible patterns behind human movements \cite{brockmann/nature/2006, gonzalez/nature/2008, song/naturePhys/2010}. Less is known about patterns behind career moves at an institutional level and how these moves affect individual performance. 

Here we take advantage of the fact that scientists publish somewhat regularly along their career \cite{petersen2012persistence, petersen2013reputation}, and for each publication, the institution in which the work was performed is listed as an affiliation in the paper, documenting career trajectories at a fine scale and in great detail. These digital traces, offering data on not only individual scientific output at each institution but also career moves from one institution to another, can provide insights for science policy, helping us understand how institutions shape knowledge, the typical moves of individual career development and help us evaluate scientific outcomes associated with professional mobility.

\begin{figure*}[!htb]
\centering
\includegraphics[width=0.8\textwidth]{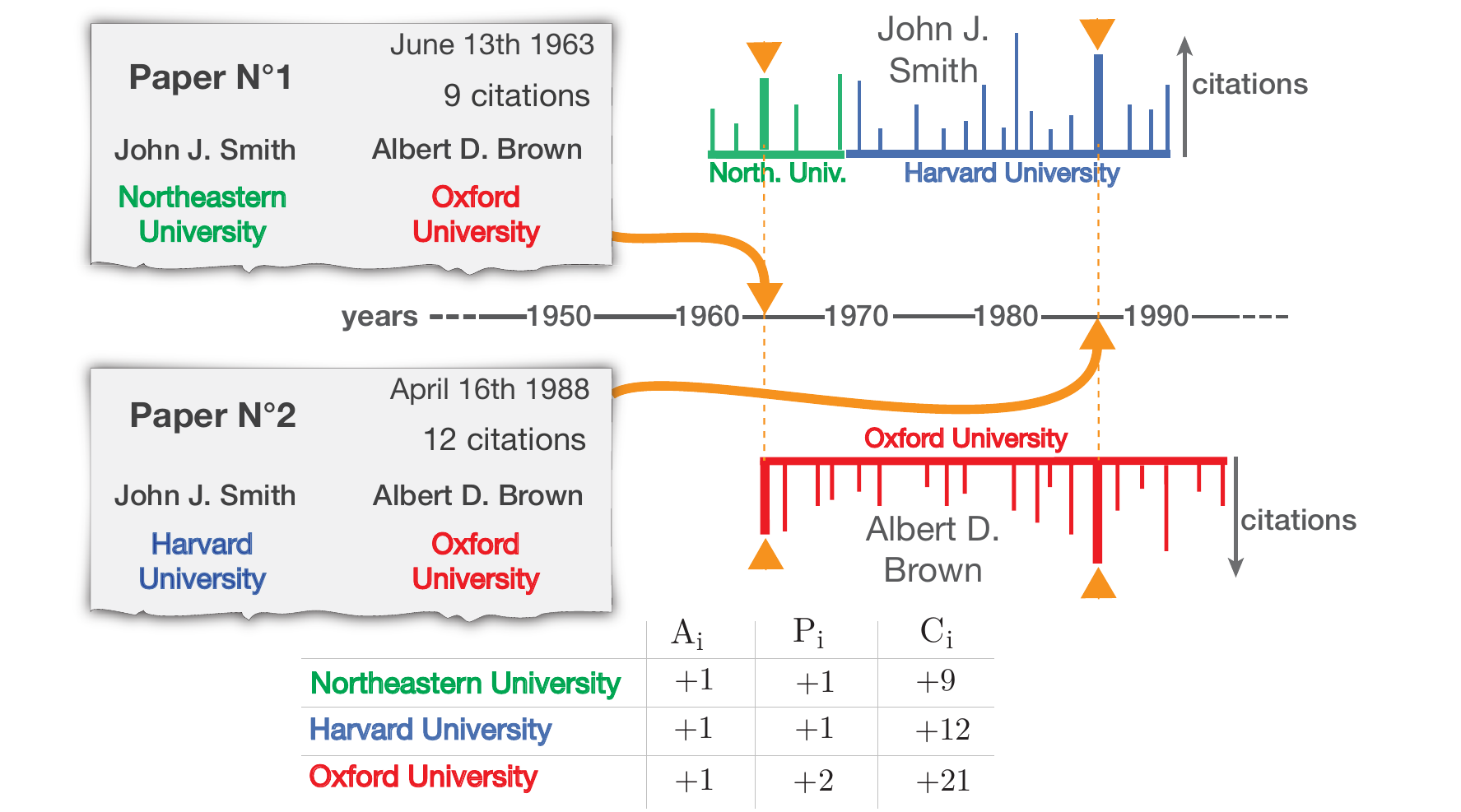}
\caption{\textbf{Illustrative example of career trajectory reconstruction for hypothetical authors.} Given the paper \textit{N\textdegree1} and \textit{N\textdegree2}, we know that the author \textit{John J. Smith} was affiliated to Northeastern University in 1963 and Harvard University in 1988. Extracting information from all his other publications allows us to reconstruct his career trajectory and discover that he was affiliated to Northeastern University for 8 years where he published 5 papers and then moved to Harvard University for 23 years where he published 16 papers. The cumulative number of citations of a paper obtained within 5 years after the publication is also known.}
\label{fig1}
\end{figure*}

We use the Physical Review dataset to extract mobility information, publication record, and citations for individual scientists. The data consists of  237,038 physicists and 425,369 scientific papers, out of which 4,052 different institutions are extracted after the disambiguation process for authors and affiliations (see SM for disambiguation process). To reconstruct the career trajectory of a scientist, we use the affiliation given in each of his/her publications (Fig \ref{fig1}). For authors with multiple affiliations listed on a paper we consider the first affiliation as primary institution. We compute the impact of each paper by counting its cumulative citations collected 5 years after its publication \cite{jones2008multi,radicchi2008universality,barabasi2012publishing,wang2013quantifying}. 

\section{Results}

\begin{figure*}[!htb]
\centering
\includegraphics[width=0.8\textwidth]{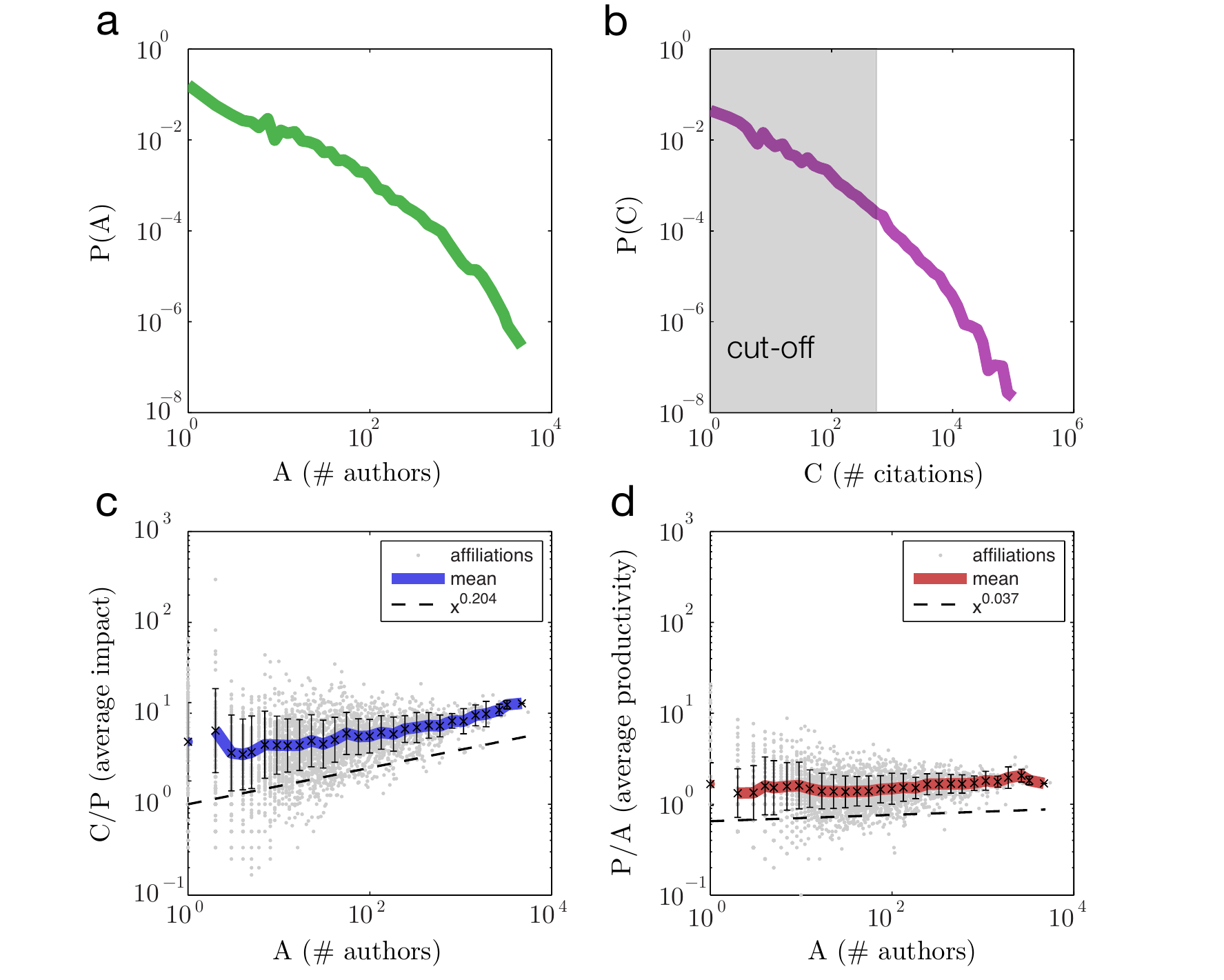}
\caption{{\bf Basic features of research institutions.} (a) The probability density function of institution size, $A$, follows a fat tailed distribution, indicating a significant heterogeneity. While most institutions size are small, a few have a large population, often representing large institutes or universities with a long history. (b) The probability density function of citations of institutions, $C$, is also very heterogeneous. Few institutions acquired a large number of citations, while most research labs or universities received few citations. Only the first thousand locations are taken into account in further analyses (shaded area). (c) The correlation between institution size and average publication impact is reported. Institution size positively correlates with the impact of publications ($R^2=0.9$) , indicating that large institutions offer a more innovative/higher impact environment than smaller ones as captured by citations per paper. The dashed line indicates a power-law behaviour with exponent $\alpha=0.204\pm0.006$ (d) The correlation between institution size and institution average productivity is also reported, indicating institution size has little influence on productivity ($R^2=0.43$). The dashed line indicates a power-law behaviour with exponent $\alpha=0.037\pm0.003$.}
\label{fig2}
\end{figure*}

Three characteristics are computed for each institution $i$ (Fig.~\ref{fig2}): the institution size ($A_i$), representing the total number of distinct authors that published at least one paper at institution $i$; the number of papers ($P_i$) published under affiliation $i$; the cumulative number of citations $C_i$ collected by all papers $P_i$. We find that $P(A)$ follows a fat tailed distribution, indicating significant population heterogeneity among different institutions (Fig.~\ref{fig2}a). While most institutions are small, a few have a large number of scientists, often corresponding to large institutes or universities. We observe similar disparity in $P(C)$ (Fig. \ref{fig2}b): few institutions acquire a large number of citations, while most research labs or universities receive few citations. 

 Figures \ref{fig2}c-d show the correlation between the institution size $A$ and both the average publications impact ${C}/{P}$ and the average productivity ${P}/{A}$ of institutions. The average productivity and impact of an institution are different but complementary measures of scientific performance. We find the institution size has little influence on productivity ($R^2=0.43$) (Fig.~\ref{fig2}d), yet it positively correlates with the impact of publications ($R^2=0.85$), indicating that large institutions offer a more innovative/higher impact environment than smaller ones as captured by citations per paper (Fig.~\ref{fig2}c). Also, as larger institutions have more internal collaborations, the number of co-authors in publications from large institutions might be larger and, as a consequence, attracts more citations \cite{jones2008multi}.
 
Many institutions are small with few citations, hence they account for very small portion of the data. For the rest of the paper, we will focus on the thousand most cited institutions, accounting for more than $99\%$ of papers. They correspond to institutions with at least 698 citations within the APS data over the 120-year period (shaded area in Fig.~\ref{fig2}).

Mobility is often important in furthering a professional career \cite{schiermeier2011career}. In science, the best lab for the type of research you are doing is usually not where you are \cite{zhang2013characterizing,borner2005spatio, mazloumian2013global}. Nowadays changing countries is a rite of passage for many young researchers who follow the resources and facilities \cite{vannoorden2004,petersen2012persistence}. As the patterns and characteristics of these migrations are blurry, we need to systematically study the mobility of scientists. Thanks to the large disambiguated data spanning the last 120 years that we have compiled, a systematic study of scientific mobility is now possible.

\begin{figure}[!htb]
\centering
\includegraphics[width=0.45\textwidth]{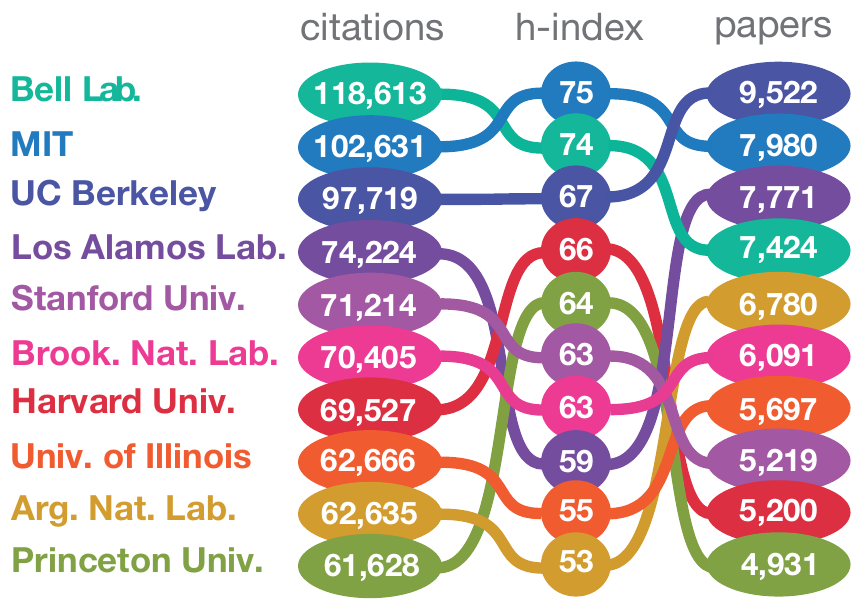}
\caption{\textbf{Ten most cited institutions in physics}. Comparison between different rankings. The H-index is closely related to the number of citations as we can observe. Top-ranked institutions all correspond to well-known universities or research labs with long tradition of excellence in physics, corroborating our hypothesis that $C$ is a reasonable proxy for ranking   }
\label{fig3}
\end{figure}

The strong correlations between the three quantities ($A, P, C$) indicate any of the three could characterise an institution, serving as a proxy of its ranking against others. Here, we choose $C$ (the total number of citations) as our parameter to approximate the ranking by reputation. Other parameters such as the h-index of an institution or the number of papers $P$ could also be used \cite{hirsch2005index,hirsch2007does,lehmann2006measures}. But the results should be insensitive to this choice owing to good correlations between these quantities ($R^2=0.96$ and $R^2=0.92$ respectively).  The top-ranked institutions all correspond to well-known universities or research labs with long tradition of excellence in physics (Fig.~\ref{fig3}), corroborating our hypothesis that $C$ is a reasonable proxy for ranking. We can also observe the similarity and stability of other rankings when comparing with other metrics.

We focus on authors with similar career longevity, restricting our corpus to those who began their career between 1950 and 1980 and published for at least 20 years without any interruption exceeding 5 years. Following these criteria, we arrived at a subset of 2,725 scientists to study the mobility patterns and their impact  on their careers. A total of 5,915 career movements are recorded for this corpus. 

\begin{figure*}[!htb]
\centering
\includegraphics[width=0.8\textwidth]{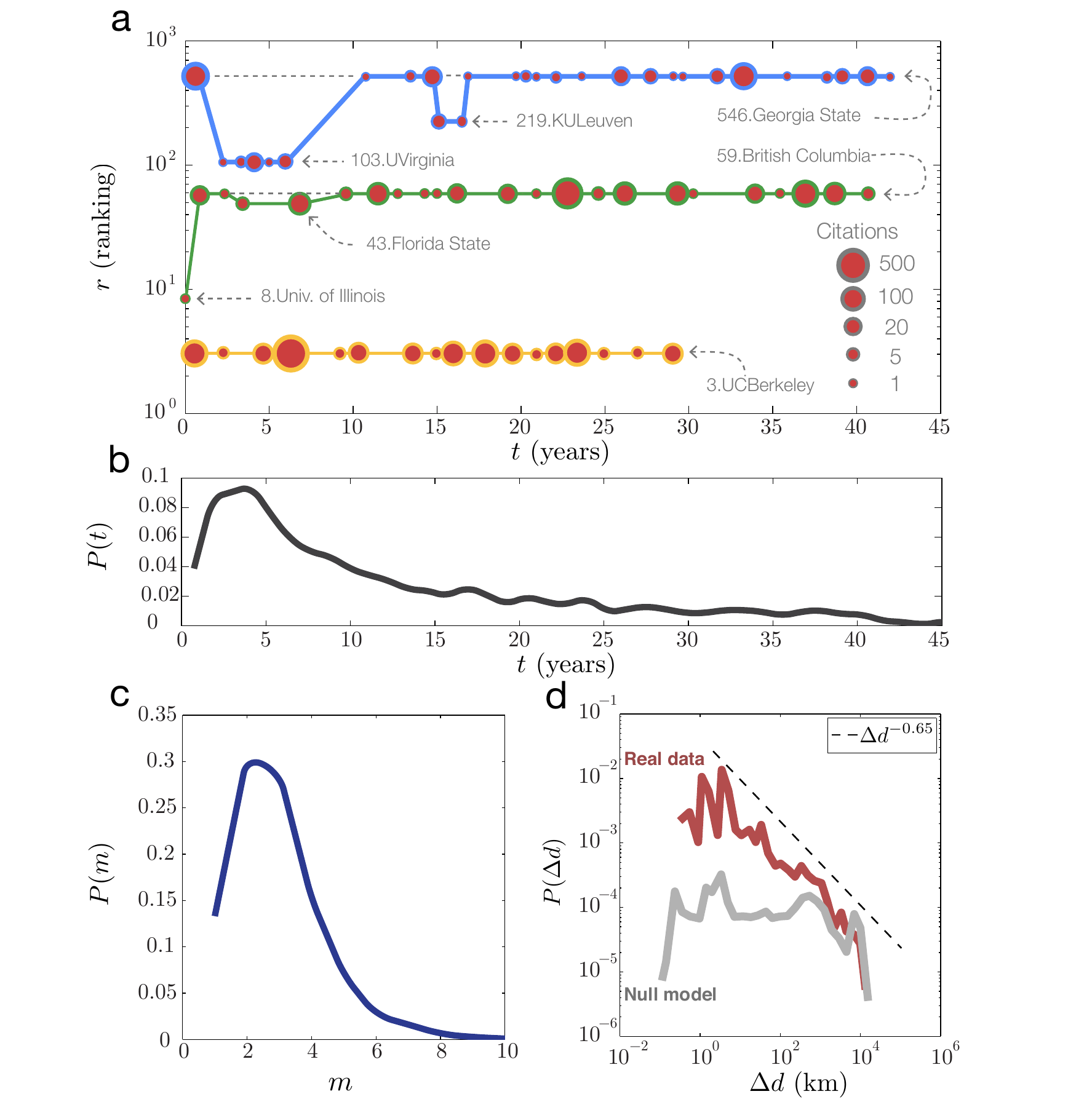}
\caption{\textbf{Basic features of scientists career.} (a) Illustration of three scientific trajectories based on publications where each line corresponds to one scientist and each publication is represented by a circle whose size is proportional to its number of citations cumulated within $5$ years after its publication. The institutions are ranked according to the total number of citations they obtained (see Methods), $1$ being the most cited institution. (b) The probability density function of movement according to time, $P(t)$, shows that most movements occurred in the early stage of the career. This likely corresponds to the postdoc period where graduates broaden their horizons through mobility. (c) The probability density function of number of visited institutions for a scientist along his career, $P(m)$, indicates that career movements are common but infrequent. Scientists mostly move once or twice, $P(m)$ decaying quickly as $m$ increases. (d) The probability density function of distance of movements, $P(\Delta d)$, has a fat-tail that can be fitted by a power law with an exponent $\gamma= 0.65\pm0.053$, whereas the null model predicts this probability to be roughly flat.}
\label{fig4}
\end{figure*}

In Figure \ref{fig4}a we select three individuals as exemplary career histories. Each line represents one individual, with circles denoting his/her publications, allowing us to observe his/her location. The size of the circle is proportional to citations the paper acquires in five years, approximating the impact of the work. By studying the whole corpus, we compute $P(m)$, the probability for a scientist to have visited $m$ different institutions along his career (Fig.~\ref{fig4}c), finding that career movements are common but infrequent: Only $14\%$ of them never moved at all ($m=1$). For the ones that move, they mostly move once or twice, $P(m)$ decaying quickly as $m$ increases. We also compute $P(t)$, the probability to observe a movement at time $t$, where $t=0$ corresponds to the date of the scientist's first publication. We find that most movements occurred in the early stage of the career (Fig.~\ref{fig4}b), supporting the hypothesis that changing affiliations is a rite of passage for young researchers \cite{schiermeier2011career}. This likely corresponds to the postdoc period where graduates broaden their horizons through mobility.  This may also reflect the increasing cost of relocation and family constraints as family developed \cite{vannoorden2004,jans2010study}. A third characteristic is the geographical distance of movements, $\Delta d$. 
Existing literature hints for somewhat competing hypothesis in the role geography plays in career movements. 
Indeed, research on human mobility suggests that regular human movements mostly cover short distances with occasional longer trips, characterized by a power law distance distribution \cite{erlander1990gravity,simini2012universal,brockmann/nature/2006,gonzalez/nature/2008}; in contrast, country-level surveys find increasing cross-country movements mostly due to cultural exposure and life quality concerns, indicating potential dominance in long distance moves in career choices comparing with typical human travels \cite{auriol2010careers,auriol2007labour,levin1999foreign,zucker2007star,vannoorden2004,franzoni2012foreign,jans2010study}. 
We measure the distance distribution over all moves observed in our dataset, finding that our result is supported by a combination of both hypothesis. 
We find the probability to move to further locations decays as a power law \cite{newman2005power,milojevic2010power} , whereas the null model predicts this probability to be flat (Fig.~\ref{fig4}d). This observation is consistent with studies on human mobility, that short distance moves dominate career choices. Yet, when comparing the power law exponents, we find the exponent characterizing career moves ($\gamma= 0.65\pm0.053$) is much smaller than those observed in human travel ($\gamma\approx 2$), corresponding to higher likelihood of observing long range movements. This observation might be explained by the influence that scientific collaborations can have on career movements as similar low exponents are observed for collaboration network between cities \cite{pan2012world}.

\begin{figure*}[!htb]
\centering
\includegraphics[width=0.8\textwidth]{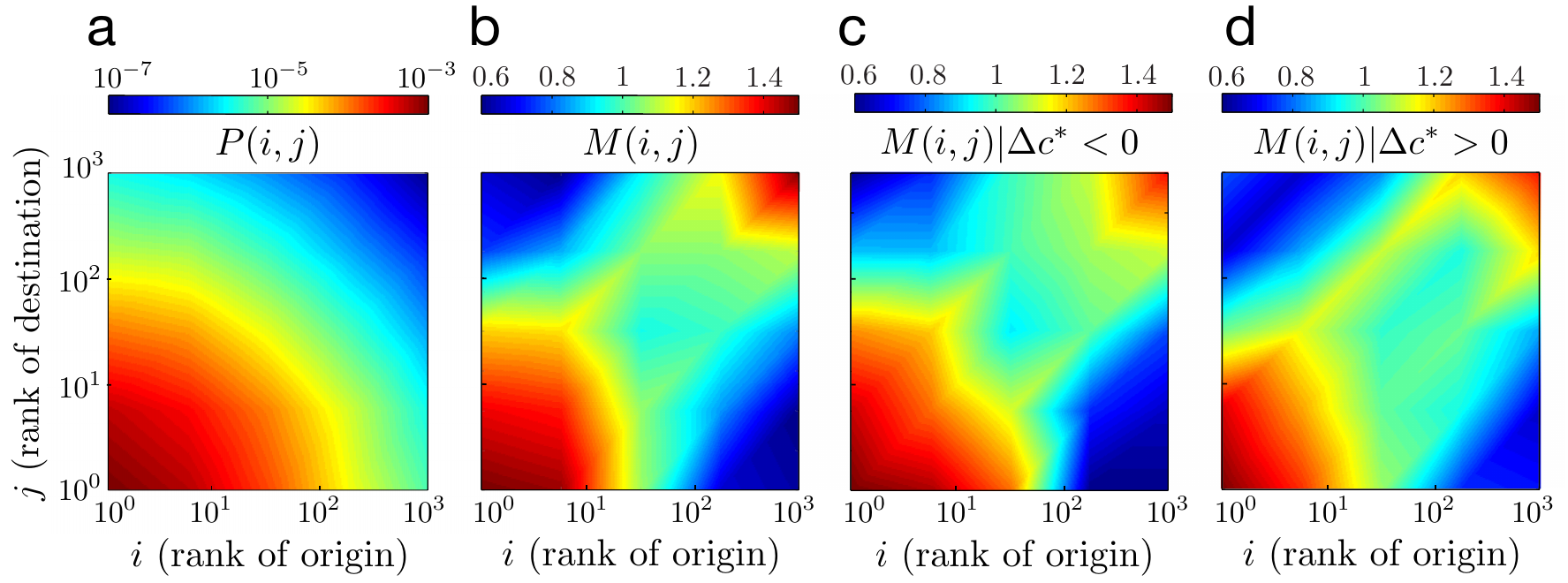}
\caption{\textbf{Stratification of career movement.} (a) The matrix of probability to have a transition from rank $i$ to rank $j$, ($1$ being the top institution) indicates that most movements involve elite institutions (rank is small) while transitions between bottom institutions are rather rare. (b) The likelihood $M(i,j)$ for a move to take place by accounting for the size of the institutions is characterized by a high degree of stratification in institutional rankings. Indeed, we observe two distinct clubs (red regions), indicating that the overrepresented movements are the ones within elite institutions (lower-left corner) or within lower-rank institutions (upper-right corner), and scientists belonging to one of the two groups tend to move to institutions within the same group. (c)-(d) The Likelihood $M(i,j)|\Delta c^*<0$ and $M(i,j)|\Delta c^*>0$ for transitions resulting in higher and lower scientific impact, respectively,  indicates that the stratification in career moves is robust against individual performance. We find the red region in lower-left corner is more concentrated in Fig.~\ref{fig5}d than in c, hinting that being more mobile in the space of rankings may lead to variable performance.}
\label{fig5}
\end{figure*}

Taken together, the preceding results indicate that career moves mostly happen during the early stage of a career and are more likely to cover short distances. The observed location in both time and space raises the question of how individual moves as a function of institutional rankings. 
To this end, denoting with $T_{i,j}$ the number of transitions from the institution of rank $i$ to the one of rank $j$, we measure $P(i,j)$, the probability to have a transition from rank $i$ to rank $j$ as
\begin{equation}\label{e1}
P(i,j)=\frac{T_{i,j}}{\sum\limits_{i,j} T_{i,j}}.
\end{equation}
Interestingly, we find that most movements involve elite institutions (rank is small), and transitions between bottom institutions are rare (Fig.~\ref{fig5}a). This is due to the fact that elite institutions are characterised by larger populations, hence translating into more events. 

To account for the population based heterogeneity, we compare the observed $P(i,j)$ with the probability $P^{null}(i,j)$ expected in a random model where we randomly shuffle the transitions from institution $i$ to $j$ while preserving the total number of transitions from and to each institution.
Formally, in this null model, we have
\begin{equation}\label{e2}
P^{null}(i,j)={\sum\limits_k P(k,j) \cdot\sum\limits_l P(i,l)},
\end{equation}
and we compare $P(i,j)$ with the null model by computing the matrix
\begin{equation}\label{e3}
M(i,j)=\frac{P(i,j)}{\sum\limits_k P(k,j) \sum\limits_l P(i,l)}.
\end{equation}
$M(i,j)$ is the ratio between the probability $P(i,j)$ to have a transition from rank $i$ to $j$ divided by the probability $P^{null}(i,j)$ when the movements are shuffled, measuring the likelihood for a move to take place by accounting for the size of the institutions. Hence, $M(i,j)=1$ indicates the amount of observed movements is about what one would expect if movements were random. Similarly, $M(i,j)>1$ indicates that we observe more transitions from $i$ to $j$ than we expected, whereas $M(i,j)<1$ corresponds to transitions that are underrepresented. We find that career moves are characterized by a high degree of stratification in institutional rankings (Fig.~\ref{fig5}b). Indeed, we observe two distinct clubs (red spots in Fig.~\ref{fig5}b), indicating that the overrepresented movements are the ones within elite institutions (lower-left corner) or within lower-rank institutions (upper-right corner), and scientists belonging to one of the two groups tend to move to institutions within the same group. 
On the other hand, both upper-left and lower-right corners are colored blue, indicating cross group movements (transitions from elite to lower-rank institutions and vice-versa) are significantly underrepresented. Also, scientists from medium-ranked institutions move to the next institution with a probability that is indistinguishable from the random case. In other words, their movements indicate no bias towards middle, elite or lower-ranked institutions.

The high intensity of stratification in career movements raises an interesting question: how does individual performance in science relate to their moves across different institutional rankings ? 

To answer this question, we need to quantify the performance change for each individual before and after the move. 
Imagine that a scientist moves from $i$ to $j$, and published $n$ papers at location $i$ and $m$ papers at $j$. 
The impact of a paper $k$ can be approximated by $c_k$, the number of citations cumulated within 5 years after its publication \cite{jones2008multi,radicchi2008universality,barabasi2012publishing,wang2013quantifying}. 
Let $c^{-}=\{c^{-}_{1},c^{-}_{2},...,c^{-}_{n}\}$ and $c^{+}=\{c^{+}_{1},c^{+}_{2},...,c^{+}_{m}\}$ be the lists of number of citations for papers published before ($c^{-}$) and after ($c^{+}$) the transition from $i$ to $j$ ($T_{i,j}$). To quantify the change in performance, we introduce 
\begin{equation}\label{e4}
\Delta c^*=\frac{\overline{c^+}-\overline{c^{-}}}{\sigma_c}
\end{equation}
where $\overline{c^+}$ and $\overline{c^-}$ are the average of $c^{+}$ and $c^{-}$, respectively, and $\sigma_c$ corresponds to the standard deviation of the concatenation of both $c^{+}$ and $c^{-}$ while preserving the moment when the movement took place (see SM for more information about $\sigma_c$). Therefore, $\Delta c^*$ captures the statistical difference in the average citations between papers published before and after the movement normalized by the random expectation when the same author's publications were shuffled.
A positive $\Delta c^*$ indicates papers following the move on average result in higher citation impact, hence representing an improvement in scientific performance. A negative value corresponds to a decline in performance.

To quantify the influence of movements on individual performance, we divide all movements into two categories based on the performance change: movements associated with positive and negative $\Delta c^*$, and measure $M(i,j|\Delta c^*>0)$ and $M(i,j|\Delta c^*<0)$. 
We find the observed stratification in career moves is robust against individual performance (Fig.~\ref{fig5}c-d). That is, the two clubs emerge for both categories in a similar fashion as in Figure \ref{fig5}b, indicating the pattern of moving within elite or lower-rank institutions is nearly universal for people whose performance is improved or decreased following the move. 
Comparing Figure \ref{fig5}c and Figure \ref{fig5}d, we find the red spot in lower-left corner is more concentrated in Figure \ref{fig5}d than in Figure \ref{fig5}c, hinting that being more mobile in the space of rankings may lead to variable performance. 
To test this hypothesis, for each transition $T_{i,j}$ we calculate the rank difference between the origin and destination ($\Delta {r_{ij}}=i-j$).  

\begin{figure}[!htb]
\centering
\includegraphics{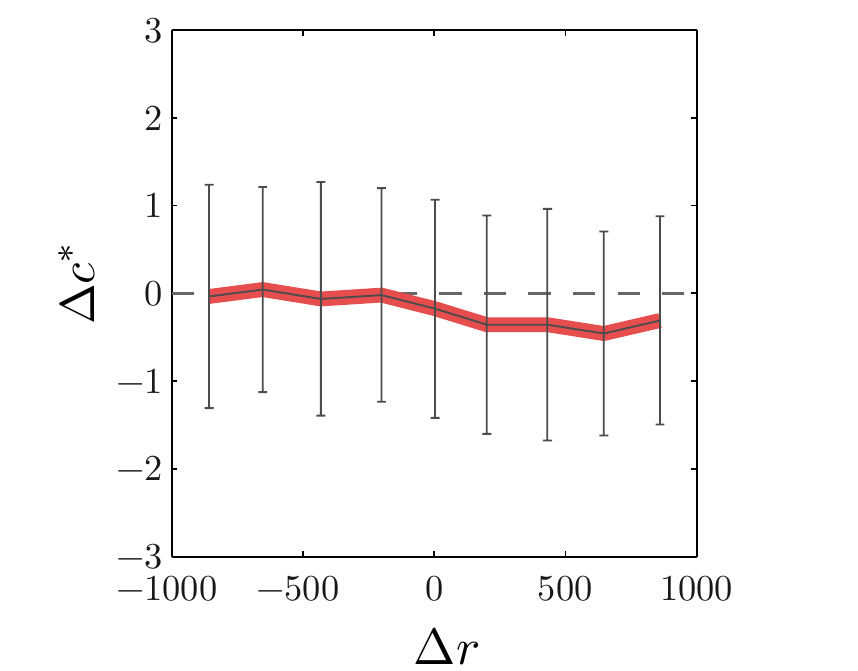}
\caption{\textbf{Impact of movements on career performance.} The relation between the statistical difference of citations ($\Delta c^*$) and the ranking difference ($\Delta {r}$) associated to a transition shows that, when people move to institutions with a lower rank ($\Delta {r}>0$), their average change in performance  is negative, corresponding to a decline in the impact of their work. Yet, what is particularly interesting lies in the $\Delta {r}<0$ regime. Indeed, when people move from lower rank location to elite institutions, we observe no performance change on average.}
\label{fig6}
\end{figure}

A positive value of $\Delta {r_{ij}}$ indicates $i>j$, hence a movement to a lower-rank institution, whereas $\Delta {r_{ij}}<0$ corresponds to transitions into institutions with a higher rank. In Figure \ref{fig6} we measure the relation between $\Delta c^*$ and $\Delta {r}$. 
When scientists move to institutions with a lower rank ($\Delta {r}>0$), we find that their average change in performance  is negative, corresponding to a decline in the impact of their work. Yet, what is particularly interesting lies in the $\Delta {r}<0$ regime. Indeed, when people move from lower rank location to elite institutions, we observe no performance change on average. This is rather unexpected, as transitioning from lower-rank institutions to elite institutions is thought to provide better access to ideas and lab resources, which in turn should fuel scientific productivity. A possible explanation may be that scientist who have the opportunity to make big jumps in the ranking space may have already had an excellent performance in their previous institutions. A move therefore will not affect their impact.

\section{Discussion}

In summary, we extracted affiliation information from the publications of each scientist, allowing us to reconstruct their career moves between different institutions as well as the body of work published at each location. We find career movements are common yet infrequent. Most people move only once or twice, and usually in the early stage of their career. 
Career movements are affected by geography. The distance covered by the move can be approximated with a power law distribution, indicating that most movements are local and moving to faraway locations is less probable. We also observe a high degree of stratification in career movements. People from elite institutions are more likely to move to other elite institutions, whereas people from lower rank institutions are more likely to move to places with similar ranks. We further confirm that the observed stratification is robust against the change in individual performance before and after the move. 
When cross-group movement occurs, we find that while going from elite to lower-rank institutions on average results in a modest decrease in scientific impact, transitioning into elite institutions, does not result in gain in impact. 

The nature of our dataset restricted our study on a sample of scientists. As a result of this selection process, our results are biased towards physicists from 1960s to 1980s with high career longevity. Yet, these limitations also suggest new avenues for further investigations. Indeed, as datasets become more comprehensive and of higher resolution, newly available data sources like Web of Science or Google Scholar can provide new and deeper insights towards generalization of the results across different disciplines, temporal trends, and more. Further investigations regarding the influence of career longevity on scientific mobility should also be considered as it could reveal as well results of importance. 
Taken together our results offer the first systematic empirical evidence on how career moves affect scientific performance and impact.

\section{Method}

{\bf Dataset.} 
The data provided by the American Physical Society (APS) contains over $450,000$ publications, each identified with a unique number, corresponding to all papers published in 9 different journals, namely Physical Review A, B, C, D, E, I, L, ST and Review of Modern Physics, spanning a period of 117 years from 1893 to 2010. For each paper the dataset includes title, date of publication (day,month,year), author names and affiliations of each of the authors. A separate dataset also provides list of citations within the APS data only, using unique paper identifiers. About 5\% of publications with ambiguous author-affiliation links or massively authored were removed from this dataset (see SM for more details).

{\bf Author Name Disambiguation.}
To derive individual information, one has to reconnect papers belonging to a single scientist. Since no unique author identifier is present in the data, author names must be disambiguated. The dataset contains about $1,2$ millions of author-paper pairs. To overcome the ambiguities present in the data, we design a procedure that uses information about the author but also metadata about the paper such as coauthors and citations. By computing similarities between authors, our procedure can successfully detect single authors as well as homonymies (see SM for more details about the disambiguation method). A total of $237,038$  distinct scientists are detected by our method.

{\bf Affiliation Disambiguation.}
A major disadvantage when dealing with publication data is the inconsistencies and errors associated with affiliation names on papers. A total of $319,829$ different affiliation names are identified in the dataset. The disambiguation procedure for affiliations uses geocoded information as well as a similarity measure between affiliation names in order to disambiguate institutions. The disambiguated set of authors also plays a crucial role in the procedure (see SM for more details about the disambiguation method). A total of $4,052$ distinct institutions are identified by our algorithm.

{\bf Resolving individual career trajectory}
Based on the information present in the publications of a scientist, we can reconstruct his/her career trajectory. In order to detect career movements, i.e. changes in a scientist's institution, one has to remove artificial movements induced by short-term stays and by errors and typos in the affiliation names on the papers. To do so, only institutions reported in at least two consecutive papers are considered in a career trajectory.

{\bf Ranking the institutions}
Three variables are considered to rank an institution: (i) the total number of papers, $P_i$, published with institution $i$,  (ii) the cumulated number of citations, $C_i$, corresponding to institution $i$, (iii) the h-index, $H_i$, of institution $i$. The variable $C_i$ is defined as $C_i=\sum\limits_{k=1}^{P_i}c_k$ where $c_k$ is the number of citations within the APS data of paper $k$ cumulated within $5$ years after its publications. An institution has an h-index $H$ if $H$ of its $P$ papers have at least $H$ citations each, and the other $(P-H)$ papers have no more than $H$ citations each. $H$ for papers  indicates the cumulative number of citations obtained within $5$ years after the publication.

{\bf Binning the institutions.}
About $6,000$ transitions between $1,000$ institutions are detected for our subset of scientists.  In order to have a statistically significant number of transitions to derive the values of $P(i,j)$ and $M(i,j)$ ( Fig.~\ref{fig5}), institutions are binned logarithmically according to their rank ($r$) into five groups.

\section{Acknowledgments}
We thank Nicolas Boumal and colleagues from the Center for Complex Network Research (CCNR) for the valuable discussions and comments.
DW, CS, and ALB are supported by Lockheed Martin Corporation (SRA 11.18.11), the Network Science Collaborative Technology Alliance is sponsored by the U.S. Army Research Laboratory under agreement W911NF-09-2-0053, Defense Advanced Research Projects Agency under agreement 11645021, and the Future and Emerging Technologies Project 317 532 "Multiplex" financed by the European Commission.

PD is supported by the National Fund for Scientific Research (FNRS) and by the Research Department of the Communaut\'e fran\c{c}aise de Belgique (\textit{Large Graph} Concerted Research Action). RS acknowledges support from the James S. McDonnell Foundation. 

\section{Author contributions}
PD designed research, analysed the data and wrote the paper. DW, RS, CS, VB and ALB, designed research and wrote the paper.

\section{Additional Information}

{\bf Competing financial interests:} The authors declare no competing financial interests.

\bibliographystyle{naturemag}


\end{document}